%% file: main.tex
\documentclass{article} %
\usepackage[preprint]{colm2026_conference}

\usepackage{microtype}
\usepackage{hyperref}
\usepackage{url}
\usepackage{lineno}

\usepackage{booktabs}
\usepackage{multirow}
\usepackage{graphicx}
\usepackage{amsmath}
\usepackage{xspace}
\usepackage{array}
\usepackage[table]{xcolor} 
\usepackage{tikz}
\usepackage{pgfplots}
\usepgfplotslibrary{groupplots}
\usepackage{subcaption}
\usepackage{tcolorbox}
\usepackage{enumitem}
\usepackage{pifont}
\usepackage{float}

\pgfplotsset{compat=newest}
\usetikzlibrary{shapes.geometric, arrows.meta, positioning, calc, fit, backgrounds}

\definecolor{darkblue}{rgb}{0, 0, 0.5}
\hypersetup{colorlinks=true, citecolor=darkblue, linkcolor=darkblue, urlcolor=darkblue}

\newtcolorbox{prompt}[1]{colback=gray!20,colframe=gray!50!black,fonttitle=\bfseries,title=#1}

\title{A Unified Model and Document Representation for\\ On-Device Retrieval-Augmented Generation}

\author{Julian Killingback\thanks{Work done while at Google.}~ \& Hamed Zamani \\
University of Massachusetts Amherst \\
\texttt{\{jkillingback,zamani\}@cs.umass.edu} \\
\AND
Ofer Meshi, Henry Li, \& Maryam Karimzadehgan \\
Google \\
\texttt{\{meshi,henryhl,maryamk\}@google.com}
}

\begin{document}

\ifcolmsubmission
\linenumbers
\fi

\maketitle
\begin{abstract}
Traditional Retrieval-Augmented Generation (RAG) approaches generally assume that retrieval and generation occur on powerful servers removed from the end user. While this reduces local hardware constraints, it introduces significant drawbacks: privacy concerns regarding data access, recurring maintenance and storage costs, increased latency, and the necessity of an internet connection. On-device RAG addresses these challenges by executing the entire pipeline locally, making it ideal for querying sensitive personal information such as financial documents, contact details, and medical history.
However, on-device deployment necessitates a delicate balance between limited memory and disk space. Specifically, the context size provided to the generative model must be restricted to manage KV cache and attention memory usage, while the size of stored embeddings must be minimized to preserve disk space. In this work, we propose a unified model that compresses the RAG context and utilizes the same representations for retrieval. This approach minimizes disk utilization compared to using separate representations, while significantly reducing the context size required for generation.
With an average of $1/10$ of the context, our model matches the performance of a traditional RAG reader without increasing storage requirements compared to a multi-vector retrieval model. This approach represents the first model to unify retrieval and context compression using a shared model and representation. We believe this work will inspire further consolidation of distinct models to optimize on-device performance.
\end{abstract}

\section{Introduction}
Retrieval-Augmented Generation (RAG) has become an indispensable part of many Artificial Intelligence (AI) workflows \citep{Lewis2020,REML}. RAG augments an original request with additional context retrieved from an information corpus, providing generative models with the external knowledge necessary to satisfy complex queries.
Existing RAG systems are typically deployed on dedicated resourceful servers for shared tasks, sometimes requiring retrieval from datasets containing trillions of tokens \citep{borgeaud2022}. While effective, this paradigm has notable drawbacks. First, it requires server access to the information corpus, raising privacy concerns for sensitive data like personal text messages, financial documents, or medical records. Second, server-based RAG relies on a constant internet connection, introducing network latency and rendering it unreliable in areas with poor coverage. Third, for personalized use cases, the cost of constructing and storing user-specific retrieval indices cannot be amortized across many users, increasing service costs and limiting accessibility.

On-device RAG addresses these issues by executing the entire pipeline locally, ensuring strict privacy, offline availability, and zero ongoing server costs. However, user devices pose strict compute and memory constraints, necessitating smaller models and minimized context sizes. To manage context limits, corpus documents can be compressed into a smaller number of vector representations \citep{500x_compressor, cocom_rag_compression, xrag_extreme_context_compression}. While creating and storing these representations requires upfront compute and disk space, this preprocessing can be opportunistically scheduled (e.g., while the device is charging), ultimately reducing the computational burden during latency-sensitive user queries.

In this work, we present a unified model for on-device RAG that performs three distinct roles: encoding information for retrieval, compressing text into vector representations, and generating responses based on those compressed contexts. Crucially, we unify the representations themselves, using the same vectors for both retrieval and generation. This eliminates the need to encode and store separate retrieval and compression representations, effectively halving the preprocessing compute, storage requirements, and memory overhead. Compared to standard RAG with a late-interaction retrieval model, our approach uses slightly less disk space while significantly outperforming standard RAG under constrained context budgets and matching standard reader performance with up to a $16\times$ compression in document representations. Compared to standard compress-and-generate RAG setups, our approach uses half the storage and significantly outperforms the context compression reader with the same context budget.

As outlined, this unification brings significant benefits, but implementing such a system is non-trivial. One approach might be to use a model and representation trained for retrieval or compression and adapt them to the other task, but this can result in catastrophic forgetting of the original task. This can be remedied by freezing the representations and using separate models, but then the representations are not optimized for one of the tasks which impacts performance. In order to get the best results, all components must be trained jointly. This is difficult, as our ablation results show. We find it requires learned scaling for the retrieval labels to ensure the retrieval information does not overwhelm the compression information. Additionally, the training data and objective must be carefully constructed to maximize the performance. We hope that our careful experimentation shows that although difficult, the unification of these skills is possible and comes with substantial benefits for on-device RAG.

Beyond on-device RAG, our strategy of unifying multiple functional roles and representations into a single model provides a blueprint for highly capable, efficient on-device ``everything-models.'' By minimizing architectural redundancy and representation storage, we hope this work demonstrates the viability of consolidated models and expedites the development of advanced on-device AI.

\input{figures/main_ecg_overview_figure}

\section{Related Works}
\noindent\textbf{Context Compression.} 
Context compression reduces memory and compute usage during generation and falls into two categories: hard and soft. 
\textbf{Hard compression} filters or rewrites text to remove irrelevant tokens. Methods like Selective Context \citep{selective_compression} and LLMLingua \citep{llm_lingua} use model probabilities to filter tokens, though this can introduce grammatical errors. Generative rewriting approaches like Nano-Capsulator \citep{nano_capsulator} and RECOMP \citep{recomp} avoid syntax issues but require expensive generation passes. Generally, hard compression is constrained by natural language bounds and often relies on unstable reinforcement learning \citep{llm_lingua_2, taco_rl, discrete_prompt_compression}.
\textbf{Soft compression} encodes context into continuous vectors. Early methods learned soft prompts \citep{prompt_compression_toxicity_reduction} or "gist" tokens \citep{gist}. Recent approaches like ICAE \citep{icae_in_context_autoencoder}, 500xCompressor \citep{500x_compressor}, COCOM \citep{cocom_rag_compression}, PISCO \citep{pisco_compression}, and ACC-RAG \citep{adaptive_context_compression} use various approaches such as auto-encoding or cross-attention to compress longer contexts for LLMs. Our approach resembles COCOM and ACC-RAG but crucially the compressed context representations are also meaningful as retrieval embeddings.
This eliminates the need for a separate retrieval model, allows for end-to-end training, and supports dynamic truncation.

\noindent\textbf{Unified Retrieval and Generation.}
\label{sec:related_work_unfied_approaches} 
Prior work largely combines only subsets of RAG tasks. Several approaches train a single model for both embedding and generation \citep{grit_lm, gem_empowering_llm_for_embedding_and_generation, onegen}, whereas xRAG \citep{xrag_extreme_context_compression} uses fixed retrieval embeddings as context for generation but maintains separate retrieval and generation models. Unlike these methods, our model unifies retrieval, compression, and generation end-to-end using shared multi-vector representations. While the concurrent work CLaRa \citep{clara_briding_retrieval_and_generation} also unifies context and retrieval representations, its evaluation is limited to reranking settings, leaving its effectiveness for full-scale corpus retrieval unknown.
Our approach is explicitly designed for the end-to-end demands of on-device RAG and offers several distinct architectural advantages. First, unlike CLaRa's fixed-length requirement, our model uses variable-length multi-vector embeddings to flexibly control retrieval and storage costs. Second, it minimizes overall disk footprint by employing a single unified model. Finally, rather than relying on frozen document vectors trained solely for compression, our method jointly trains representations for both retrieval and generation contexts.

\noindent\textbf{On-Device LLMs and RAG.} Deploying LLMs locally relies heavily on hardware-level optimizations, such as intelligently distributing and loading parameters and optimizing computation graphs \citep{llm_in_a_flash, mmn_llm, power_infer}, while more general optimizations like speculative decoding and quantization are also used \citep{llm_cad, genai_at_the_edge, accelerating_mobile_lms_speculative_decoding}. On-device RAG introduces the added challenge of managing a local index, which prior works address through compartmentalized indexes \citep{mobile_rag}, lexical filtering \citep{pocket_rag}, and improved memory management \citep{dirc_rag_edge_rag}. Our approach is largely orthogonal to these low-level optimizations. By significantly reducing context size through compressed document representations, our method can be seamlessly combined with existing techniques. Furthermore, unifying retrieval and generation representations saves disk space and reduces latency, as relevant documents are already loaded into memory immediately following the retrieval stage.

\section{Unifying Embedding, Compression, and Generation}
We propose a framework for unifying embedding, compression, and generation into a single model. An overview of using this unified model for RAG is shown in Figure \ref{fig:inference_usage_of_ecg_model}. We refer to models from our framework as ECG models denoting their three unified capabilities. This section details the architecture and multi-task training recipe used to achieve this unification.

\subsection{Retrieval Similarity} \label{sec:similarity}
\newcommand{\repq}{\mathbf{E}_{q}\xspace}
\newcommand{\repd}{\mathbf{E}_{d}\xspace}
Our query and document representations can contain multiple embeddings which enable high-fidelity compression of information, thus we need a fast similarity measure suited for multi-vector representations. With these goals in mind, we adopt a mean-pooled variant of the MaxSim similarity introduced by ColBERT \citep{colbert}: $S(\repq, \repd) = \frac{1}{m} \sum_{i} \max_{j} (\repq^i \cdot \repd^j)$. We use the mean over the $m$ query embeddings rather than a strict sum to account for variable-length query representations without penalizing queries with fewer embeddings.

\subsection{Model Representations} \label{sec:model_details}
\newcommand{\lmhidden}{\mathbf{E}\xspace}
\newcommand{\repcontext}{\mathbf{E}_{comp}\xspace}
\newcommand{\repret}{\mathbf{E}_{ret}\xspace}

\newcommand{\embstart}{s_{start}\xspace}
\newcommand{\emb}{s_{emb}\xspace}
\newcommand{\embend}{s_{end}\xspace}

To adapt a pretrained decoder transformer $\theta_{LM}$ to also act as an encoder (produce continuous embeddings), we prepend a task prompt $p_{embed}$ to the input text $t$ and append a sequence of $n + 2$ special tokens: $(\embstart, \emb, \dots, \emb, \embend)$. The final hidden states corresponding to the $\emb$ tokens, denoted $\lmhidden \in \mathbb{R}^{n \times d}$, are passed through two projection blocks. 

The first block, $\theta_{ret}: \{\lmhidden\} \to \repret \in \mathbb{R}^{n \times m}$, produces retrieval representations optimized for MaxSim similarity. Note, the dimension $m$ can differ from $d$ for efficiency, but we keep $m=d$ in all our experiments. The second block, $\theta_{comp}: \{\repret\} \to \repcontext \in \mathbb{R}^{n \times d}$, transforms $\repret$ into compressed context representations suitable as input to $\theta_{LM}$ for generation. Constructing $\repcontext$ directly from $\repret$ ensures that only one set of representations needs to be stored for both retrieval and generation. Implementation details for these blocks are provided in Section \ref{sec:experiments_model_details}.

To generate text conditioned on $k$ distinct compressed contexts, we construct the input sequence by wrapping each context representation $\repcontext^{(i)}$ in special tokens: $\text{Input} = \text{prefix} \oplus \left( \bigoplus_{i=1}^{k}[ \embstart \oplus \repcontext^{(i)} \oplus \embend ] \right) \oplus \text{suffix}$, where $\oplus$ denotes sequence concatenation.

\subsection{Unified Training Recipe}
We employ a multi-task training approach consisting of a self-supervised pretraining phase to adapt the model to continuous representations, followed by RAG-specific fine-tuning.

\paragraph{Self-Supervised Training}
In this stage, we utilize an unlabeled passage collection to teach the model to compress and understand context representations and produce meaningful retrieval representations. For a given passage, we split the text in half to create context and target pairs using two distinct strategies with equal probability: (1) \textit{reconstruction}, where the target text is identical to the context text, and (2) \textit{neighboring text}, where the target is the other half of the passage. These strategies, inspired by \citet{cocom_rag_compression}, provide complementary signals: reconstruction forces the model to exactly compress information while neighboring text requires the compressed context to be useful for predicting surrounding text. 

The context text is encoded into retrieval representations $\repret$ and compressed context representations $\repcontext$. The compressed context $\repcontext$ is then fed into $\theta_{LM}$ to generate the target text via standard next-token prediction. Simultaneously, we use $\repret$ to compute an in-batch contrastive loss, ensuring that representations for the context and target pairs have higher MaxSim similarity than representations of other random texts in the batch. 

To maximize computational efficiency, we perform generation bi-directionally (using the target as context to generate the original context) so that every encoding step provides a training signal. Furthermore, we uniformly vary the number of $\emb$ tokens during this stage, forcing the model to learn effective compression across a dynamic number of representations. The full mathematical formulation of the self-supervised losses and further discussion on how this extends prior work can be found in Appendix \ref{sec:appendix_self_supervised_loss} while the full training details can be found in Section \ref{sec:training_details}.

\paragraph{RAG Specific Training}
To adapt the model for RAG, we fine-tune it to generate answers from multiple compressed context documents. To prevent the generative component from overfitting, we replace standard next-token prediction with knowledge distillation. Specifically, we utilize two distinct teacher models: a generative reader $\theta_{reader}$ for answer distillation, and a teacher scoring model to provide query-document relevance scores.

Given a question, an answer, a positive document, hard negatives, and teacher relevance scores, we encode all queries and documents into retrieval representations $\repret$ and compressed contexts $\repcontext$ using a fixed number of $\emb$ tokens, $t$. The model is trained jointly to optimize generation and retrieval. For generation, we minimize the KL divergence between our model (conditioned on compressed contexts) and the teacher reader (conditioned on uncompressed documents). For retrieval, we apply an InfoNCE contrastive loss \citep{infonce_loss}  alongside a Margin MSE loss \citep{margin_mse} to distill the ranking capability of the teacher scoring model into our student model. The final joint loss is $\mathcal{L} = \mathcal{L}_{gen} + \mathcal{L}_{contrastive} + \mathcal{L}_{margin}$. Full mathematical formulations of these objectives are provided in Appendix \ref{sec:appendix_rag_loss_equations}, and training parameters are detailed in Section \ref{sec:training_details}.

\section{Experiment Details}
\subsection{Model Implementation} \label{sec:experiments_model_details}
We evaluate our approach using the instruction-tuned version of two pre-trained decoder language models chosen to represent sizes plausible for on-device RAG applications: SmolLM-v2 135M \citep{smollm_2} and Gemma 3 1B \citep{gemma3}. We augment the model and tokenizer with special tokens representing the start, body, and end of the embedding sequence, denoted as \texttt{<emb\_start>}, \texttt{<emb>}, and \texttt{<emb\_stop>} respectively. The projection blocks $\theta_{comp}$ and $\theta_{ret}$ share the same architecture designed to transform the hidden states effectively while maintaining training stability using a gated residual connection and linear layers. The full architectural details for $\theta_{comp}$ and $\theta_{ret}$ are in Appendix \ref{sec:appendix_theta_comp_theta_ret}.

\subsection{Training Details}
\label{sec:training_details}
\paragraph{Self-Supervised Training Details}
For the self-supervised pre-training stage, we utilize a chunked version of Wikipedia released by \citet{dense_passage_retrieval}. Each chunked passage contains approximately 100 words. We train for a single epoch, using the last checkpoint for our evaluation. For each model trained with the self-supervised pretraining stage, including ablation models, we do a hyperparameter grid search to ensure results are comparable. The hyperparameter search and selected hyperparmeters are detailed in Appendix \ref{sec:appendix_hyperparameters}.

\paragraph{RAG Training Details}
Following the self-supervised stage, we do specialized RAG training, utilizing data from the Natural Questions (NQ) \citep{natural_questions_benchmark} and Trivia QA \citep{trivia_qa} training datasets. More training data details are in Appendix \ref{sec:appendix_rag_training_details}.

During training, we sample one positive document, two negative documents, and one answer to act as a single example. A batch is constructed with several such examples. The positive document and the answer are picked randomly with equal probability, while the negative documents use a weighted sampling approach detailed in Appendix \ref{sec:appendix_rag_weighted_negative_sampling}. We found compared to sampling negatives uniformly the weighted approach had better results; we explore the impact in our ablation study in Appendix \ref{sec:appendix_ecg_model_ablations}.
The number of $\emb$ tokens to use is a constant value $t$ for both queries and documents, for the SmolLM variants we use $t=32$ while for the Gemma variants we use $t=16$. We selected these values to make the total information stored roughly the same as what a standard ColBERT model stores. Furthermore, as the hidden size of Gemma is $1,152$ and SmolLM is $576$ this means the total space for the representations is equivalent allowing for a more direct comparison between the two models. Additional training details are presented in Appendix \ref{sec:appendix_rag_training_details}.

\subsection{Evaluation}
To evaluate our method, we do end-to-end evaluation using the Exact Match (EM) accuracy metric. Although our focus is on-device RAG, due to the lack of realistic datasets for the on-device setting, such as personal document search, we use a standard set of RAG benchmarks. This has the additional benefit of allowing for easier comparison with existing work. Although we are using standard datasets, we wanted to make sure we considered the limitations that come with on-device RAG. Specifically, as context length is limited in on-device settings, our evaluation highlights performance at given context budgets. We define a context budget as the number of document representations, either representing tokens or compressed vectors, that are used by the generator. For standard reader models we truncate the document texts to ensure the context budget is met. To get a comprehensive understanding of performance-context-budget trade-off we focus on two evaluation scenarios:
\begin{description}[leftmargin=8pt, labelindent=0pt]
    \item[Fixed Context Budget] In this setup, we use a fixed number of document representations for our method and the baseline methods. This shows the relative capacity of the various methods with a constrained context budget.
    
    \item[Fixed Performance] In this setup, we use the performance of the ECG models, with a single document, as a fixed performance target and increase the baseline context budget until they match the ECG models' performance (or perform best if no context budget matches our performance). We do a grid-search from 32 to 256 in intervals of 32. This evaluation approach shows the compression that the ECG models can achieve while matching or exceeding the baseline performance.
\end{description}

The context budget is applied only to the document representations in the model's input and does not include question tokens or formatting tokens, including the special tokens $\embstart$ and $\embend$. In validation tests we found that the ECG models generally performed best with only the top ranking document and adding additional documents generally reduced performance, thus in our main evaluation we only use a single document as input to our models unless otherwise specified, we discuss this finding in more detail in Appendix \ref{sec:appendix_accuracy_drop_off}.

\subsubsection{Evaluation Datasets}
\label{sec:evaluation_datasets}
We evaluate our approach on two common RAG datasets NQ Test \citep{natural_questions_benchmark} and Trivia QA Test \citep{trivia_qa}. We use the versions released by FlashRAG \citep{flashrag}. For each dataset we use the Wikipedia passages released by \citep{dense_passage_retrieval} to form a corpus. To enable efficient evaluation we use a pooling approach to reduce the large corpus size to a more manageable size. The details can be found in Appendix \ref{sec:appendix_evaluation_corpus_details}.

\subsubsection{Validation}
To select the best checkpoints for both our model and the baseline models as well as understand our model's performance we use two validation sets. The first is 1k questions sampled from the training set described in Section \ref{sec:training_details} (the validation data is removed from the training data), we use the validation loss from this data to select the best checkpoint for the ECG and baseline models. For model development we use 500 questions sampled from the NQ validation set with a pooled set of documents to form the corpus.

\subsubsection{Baselines}
\label{sec:baselines}

For baselines, we include three alternatives to our approach:
\begin{description}[leftmargin=8pt, labelindent=0pt]
    \item[Parametric Knowledge Only] This baseline leverages the model's parametric knowledge to answer a question directly without any additional information from retrieved documents. 
    
    \item[Standard RAG] This baseline uses two models: a retrieval model and reader model. The reader model takes the question and retrieved documents as text to answer the question. We use three different retrieval models outlined below.
    
    \item[RAG with Context Compression] Like Standard RAG, this baseline uses a retrieval model, but instead of using the raw text for documents it uses compressed documents as context. To enable a direct comparison, we use the same number of representations as the ECG models. Our context compression approach is based on COCOM \citep{cocom_rag_compression} which directly uses the transformer's final hidden state as the compressed representation. We do not include xRAG as a baseline because as shown by \citet{cocom_rag_compression} it is far worse than the end-to-end multi-vector representation we adopt. Our choice of a context compression baseline is intended to serve as a reasonable proxy for strong context compression models while removing many confounding factors such as base model, training data, and modeling approach.
    \end{description}

To ensure that the comparison with the baseline models are fair, we initialize from the same starting models and use the same training data when applicable. The full training details for the baselines are in Appendix \ref{sec:appendix_baseline_training_details}. For the RAG-based methods we consider three retrieval methods that vary in effectiveness and efficiency. These are:

\begin{description}[leftmargin=8pt, labelindent=0pt]
    \item[GTE ModernBERT Base \citep{gte_models}] A dense retrieval model based on ModernBERT \citep{modern_bert} a modernized encoder model with 149M parameters. It is trained on a large scale dataset of text pairs across many domains and before being trained on high-quality retrieval datasets.
    
    \item[GTE ModernColBERT V1 \citep{gte_moderncolbert_v1}] This model is similar to the original ColBERT model, but uses GTE ModernBERT Base as the base model. It is trained on the MSMARCO passage dataset and represents one of the strongest retrieval models for its size.
    
    \item[BM25 \citep{bm25}] A classic lexical approach which does not have any learned weights, but has been shown to be highly capable and robust across domains.
\end{description}
Additional implementation details for the baseline retrieval approaches are presented in Appendix \ref{sec:appendix_implementation_and_resources}.

\section{Results and Discussion}
\label{sec:results}
\input{tables/main_results_fixed_context_budget}
To demonstrate the viability of our unified ECG model for on-device applications, we structure our evaluation around four primary research questions (RQs). On-device deployment involves strict hardware constraints; for instance, typical mobile devices operate with shared, limited unified memory. Large text contexts in standard RAG balloon the KV cache, leading to out-of-memory errors or additional battery drain, while storing multiple retrieval and context representations exhausts precious disk space. Our evaluation addresses these constraints through the following RQs:
\begin{itemize}[leftmargin=*, labelindent=16pt]
    \item \textbf{RQ1:} How does the unified ECG model perform under constrained on-device context budgets compared to the baselines including standard RAG?
    \item \textbf{RQ2:} What level of context efficiency and disk space reduction does unifying representations achieve while matching standard reader performance?
    \item \textbf{RQ3:} Does joint training of retrieval and compression representations outperform isolated context compression?
    \item \textbf{RQ4:} What factors, including architectural and training components, drive ECG's performance?
\end{itemize}

\paragraph{RQ1: Performance Under Constrained Context Budgets}
Table \ref{tab:main_results_fixed_context_budget} presents the fixed context budget evaluation, demonstrating that ECG models substantially outperform all baselines under strict active memory (KV cache) constraints. When restricted to a budget of 32 representations for SmolLM and 16 for Gemma, standard RAG models struggle significantly. In contrast, the ECG models achieve over three times the Exact Match score of the best standard RAG models on NQ for both architectures (0.343 vs. 0.106 for SmolLM; 0.361 vs. 0.104 for Gemma). While the context compression baselines fare slightly better than standard RAG at these budgets, ECG still outperforms the strongest isolated context compression baseline by substantial margins across all datasets and model sizes. Ultimately, when active memory is strictly capped, standard RAG fails, whereas ECG remains robust.

\paragraph{RQ2: Context Efficiency and Disk Space}

\input{tables/main_results_fixed_performance}
Table \ref{tab:main_fixed_performance} illustrates the context budget required for baselines to match the single-document performance of the ECG models. Standard compress-and-generate RAG systems typically require storing two distinct sets of vectors: one for retrieval (e.g., ColBERT) and one for context generation (e.g., COCOM). By sharing representations, ECG cuts disk space requirement in half, using slightly less space than ColBERT alone, while simultaneously solving the context-length memory issue. 

The results show that the vast majority of baselines---including all parametric, BM25, Dense RAG, and Context Compression models---fail to reach ECG's performance entirely, even when given up to $8\times$ or $16\times$ the context budget. Standard RAG with ColBERT retrieval is the only baseline capable of matching ECG's accuracy, but it requires a substantial context increase to do so. Specifically, it requires $5\times$ the budget on SmolLM TriviaQA, $14\times$ the budget on Gemma NQ, and $8\times$ the budget on Gemma TriviaQA. Furthermore, on the SmolLM NQ dataset, even when given a maximum budget of 256 representations ($8\times$ the ECG budget), RAG ColBERT (0.328) still falls short of the ECG model's performance (0.343). 

\paragraph{RQ3: Unified Modeling vs. Isolated Compression}
As Tables \ref{tab:main_results_fixed_context_budget} and  \ref{tab:main_fixed_performance} show, the ECG models perform better than the context compression baselines across the board. We investigate the underlying cause of this difference by training and evaluating variants which isolate various aspects of each model. Specifically, we train a context compression model with the exact same architecture as the ECG model which includes the MLP blocks and use the ECG model on ColBERT's retrievals to isolate the compression and generation quality. The results in Table \ref{tab:context_compression_same_arch} shows the performance discrepancy is partially due to the architectural differences and retrieval quality, but there is still a substantial gap when these are accounted for (0.224 vs. 0.297 on NQ and 0.386 vs. 0.485 on Trivia QA) which suggests unifying retrieval and compression not only improves efficiency, it also improves effectiveness. We hypothesize that ECG's multi-task training objective---specifically forcing the vectors to satisfy the contrastive retrieval loss---acts as regularization. This yields richer, more robust context representations than compression training alone.

\paragraph{RQ4: Driver of Performance and Architectural Impact}
The ECG models perform impressively compared to standard RAG as discussed in RQ2. In fact, in several instances standard RAG approaches cannot match the ECG models even with the maximum context. To understand what is behind this strong performance we investigate hybrid evaluations where the ECG model is used only as a retriever or only as a reader and compare against the baseline readers and retrievers. The NQ results are presented in Figure \ref{fig:nq_k_scaling}. We can see that the ECG model's performance is bolstered by excellent retrieval quality with standard readers having a clear increase when using the ECG retrieval over the best baseline retriever ColBERT. Although the ECG reader is generally worse than the standard reader, the gap is generally minimal considering the context size used by the ECG reader is far smaller. We also note that the standard reader also does best with only one or two documents when using the ECG retrievals showing that the preference for fewer documents is not unique to the ECG models. For additional discussion on this finding see Appendix \ref{sec:appendix_accuracy_drop_off}.

As we've discussed unifying retrieval, compression, and generation into a single representation space is not trivial. To understand these challenges, Table \ref{tab:ecg_ablations} presents an ablation study of the SmolLM ECG model. We find that contrastive loss during pretraining is a necessary addition for performance on NQ with a notable drop from 0.343 to 0.337 when removed, while there is minimal decrease in Trivia QA. Generative distillation is also crucial, providing substantially higher quality outputs compared to standard next-token prediction. However, the most significant challenge lies in balancing the retrieval and generation objectives during joint training. Removing dynamic loss scaling---the learned temperature and teacher score scale for the retrieval losses---causes NQ performance to plummet by nearly half (from 0.343 to 0.173). This demonstrates that without precise alignment of the teacher's scores to the student distribution, the multi-task objectives actively compete and degrade the shared representations. An expanded ablation section is available in Appendix \ref{sec:appendix_ecg_model_ablations}.

\input{figures/comparison_ecg_reader_retrieval}

\section{Conclusion}
In this work we show that retrieval and context representations for a document can be shared for RAG and that this can be done with a unified model acting as the compressor, encoder, and generator. Our approach uses the same space as other powerful late-interaction models such as ColBERT, but uses far fewer context representations for comparable performance when compared to standard text reader models. When compared with a stand-alone context compressor paired with a retrieval model, our method substantially outperforms the context compression model even when evaluated with the same context documents, indicating that combining retrieval and compression representations brings additional robustness and performance gain. Furthermore, our approach requires half the storage compared to the standard context compression approach which requires two document representations, one to act as context generation and one as the retrieval representation. Our work enables on-device RAG to have high performance with a relatively tiny context size and minimal disk usage. We hope it inspires novel and more efficient unified approaches to RAG.

\section*{Acknowledgments}
We used LLMs in various capacities throughout the experiment process and in the writing process. Specifically LLMs were used to: write code for some of the experimental setup, produce tables and figures (but not add or create their content), suggest potentially relevant related work (we never directly use an LLM to write or generate citations, we only use it to suggest papers that are manually reviewed), proofread various writing, and rewrite certain sections of the paper. All LLM outputs were carefully verified by one or more of the authors to ensure correctness. We take full responsibility for all content and findings reported in this paper.

This work was supported in part by the Center for Intelligent Information Retrieval and in part by the NSF Graduate Research Fellowships Program (GRFP) Award \#1938059. Any opinions, findings and conclusions or recommendations expressed in this material are those of the authors and do not necessarily reflect those of the sponsor.

\bibliographystyle{colm2026_conference}
\bibliography{main}

\appendix
\section{Modeling Details}
\subsection{Implementation of $\theta_{comp}$ and $\theta_{ret}$}
\label{sec:appendix_theta_comp_theta_ret}
The projection blocks $\theta_{comp}$ and $\theta_{ret}$ share the same architecture designed to transform the hidden states effectively while maintaining training stability. Both blocks consist of a stack of $L=4$ identical layers with hidden dimension $d$ matching the underlying language model. Formally, let $h_l$ denote the input to the $l$-th layer. The output $h_{l+1}$ is computed via a gated residual connection:
$$
h_{l+1} = \sigma(W_{gate} h_l + b_{gate}) \odot h_l + \text{ReLU}(W_{proj} \text{LN}(h_l) + b_{proj})
$$
where $\sigma$ is the sigmoid function, $\odot$ denotes element-wise multiplication, $W_{gate} \in \mathbb{R}^{1 \times d}$ represents the weight for the gated residual, $W_{proj} \in \mathbb{R}^{d \times d}$ represents the weight for the linear layer, and $\text{LN}$ represents Layer Normalization. For the final layer of each block we do not apply the ReLU activation so as not to restrict the representations to only positive values. To facilitate gradient flow and preserve information from the pre-trained model during the early stages of training, the bias $b_{gate}$ is initialized to a large positive value, ensuring the gate output is initially close to 1. 

The final value of $\repret$ is found by dividing by $\sqrt{m}$ where $m$ is the dimension of the output from $\theta_{ret}$. Before passing this into $\theta_{comp}$ we multiply by $\sqrt{m}$. This is done to normalize the inner products produced by $\repret$ similar to what is done during multi-head attention \citep{attention_is_all_you_need}. A more common approach in retrieval models is to normalize vectors to unit lengths, but in our context this may reduce information and impact the generation input. It would be possible to store the magnitude and rescale in the same way we do with $\sqrt{m}$, but this adds additional complexity so for simplicity we use $\sqrt{m}$.

\section{Additional Implementation, Evaluation, and Training Details}
\subsection{Implementation and Resources}
\label{sec:appendix_implementation_and_resources}
We implemented our models and performed training using PyTorch \citep{pytorch} and Huggingface Transformers library \citep{huggingface_transformers}. We used 8 NVIDIA H100 GPUs for training tasks and NVIDIA A100s for inference. For GTE ModernColBERT V1 we use the Pylate \citep{pylate} library implementation. For BM25 we use Pyserini \citep{pyserini} with default arguments.  

We perform retrieval by doing exact similarity computations for each document in the corpus (note that we use pooling to reduce the corpus sizes as mentioned in Section \ref{sec:evaluation_datasets}). This was done using custom PyTorch code. We choose this over an approximate approach because it reduces the confounding factors and because we found that existing implementations for approximate late-interaction retrieval such as PLAID were not designed for large embedding dimensions (e.g. $> 256$). Note, that in principle the algorithms are dimension agnostic it is just certain choices made in current implementations which cause problems.

\subsection{Evaluation Corpus Details}
\label{sec:appendix_evaluation_corpus_details}
To enable efficient evaluation we use a pooling approach to reduce the large corpus size of the Wikipedia passage corpus to a more manageable size. Our pooling procedure uses four different retrieval models to find the top 100 passages for each test question and combines the retrieved documents into a unified corpus. We use the three baseline retrieval models described in Section \ref{sec:baselines} and an early version of our model to construct the pooled corpora. Though this changes the retrieval task, prior work has shown pooling can be an effective way to reduce the corpus size while still ensuring a valid evaluation of full-scale retrieval performance \citep{pooling_corpus}. After the pooling process, the corpus constructed for NQ has approximately 840k passages and the corpus constructed for Trivia QA has approximately 2.1M passages.

\subsection{Baseline Training Details}
\label{sec:appendix_baseline_training_details}
To ensure that the comparison with the baseline models are fair we initialize from the same starting models and use the same training data when applicable. For the Parametric and Standard RAG reader models we train on only the RAG training data, though the parametric model only sees the question without retrieved documents. For the context compression approach we include a pretraining phase like the ECG models. The pretraining phase is identical to the ECG models except there is no contrastive loss. For the RAG training phase the data and batch processing is the same as the ECG models, but without any retrieval losses. Like the ECG models the context compression models use knowledge distillation as the main generative loss. The parametric models and reader models use next-token prediction. To ensure that hyperparameter selection is not responsible for any performance disparities, we perform a learning rate and weight decay sweep for all the baseline models. The full sweep information and final hyperparameters are detailed in Appendix \ref{sec:appendix_hyperparameters}.

\subsection{Self-Supervised Training Details and Loss Equations}
\label{sec:appendix_self_supervised_loss}
During the self-supervised training stage, our model learns to compress and leverage compressed representations by predicting target text $T$ from compressed context text. Our approach builds upon the first stage of training in COCOM \citep{cocom_rag_compression}, but with key modifications to support both compression and retrieval representation learning.

\subsubsection{Generation Loss}
Unlike COCOM, which uses a full passage for reconstruction and half-passages for the next-segment generation task, we consistently split passages in half. This ensures there is no difference in the distribution while encoding for either the reconstruction or neighboring text strategies. The generative component uses a standard next-token prediction loss over the target text tokens, conditioned on the compressed context $\repcontext$:
$$
\mathcal{L}_{lm} = - \sum_{i=1}^{|T|} \log p(t_i \mid t_{< i}, \repcontext)
$$
The loss is applied exclusively to the target text tokens, ignoring formatting tokens and the $\repcontext$ vectors.

\subsubsection{Contrastive Retrieval Loss}
To simultaneously train $\repret$ to produce meaningful representations for retrieval (using the similarity function defined in Section \ref{sec:similarity}), we apply an in-batch contrastive loss. We treat the $\repret$ of the context text ($\repret^c$) and the target text ($\repret^t$) as a positive pair. 

To prevent computational waste when encoding the target text solely for the retrieval loss, we perform generation in both directions (context $\rightarrow$ target, and target $\rightarrow$ context). Given a batch of $B$ positive pairs, the contrastive loss is computed as:
$$
\mathcal{L}_{contrastive} = - \frac{1}{B} \sum_{i=1}^{B} \log \frac{\exp(S(\repret^{c, i}, \repret^{t, i}) / \tau)}{\sum_{j=1}^{B} \exp(S(\repret^{c, i}, \repret^{t, j}) / \tau)}
$$
where $\tau$ is a learnable temperature parameter. 

Note that for the reconstruction task, the model encodes the exact same text twice. However, because we randomly vary the number of $\emb$ tokens for each encoding pass, the model effectively learns to map the same information to similar representations across different capacity constraints. This acts as a form of augmentation, allowing identical text to be used in contrastive training (similar to how standard dropout acts as augmentation in models like SimCSE \citep{sim_cse_simple_contrastive_learning}).

The final joint loss for the self-supervised stage is the sum of the generative and contrastive components:
$$
\mathcal{L} = \mathcal{L}_{contrastive} + \mathcal{L}_{lm}
$$

\subsection{Additional RAG Training Details}
\label{sec:appendix_rag_training_details}

During RAG training, the order of the context documents provided to the generative model is randomized to prevent the model from learning any positional bias. All documents are used in the retrieval losses, but to make the generative component of the model more robust, we randomize the number of documents provided for generation from one to three while ensuring that the positive document is always included. 

As the generative teacher model for the knowledge distillation, we employ the reader model trained as a baseline; further details on the teacher model can be found in Section \ref{sec:baselines}. Like in the pretraining stage we do a hyperparameter search for our models using the validation set to select the best checkpoint. The full training details including hyperparameter search details can be found in Appendix \ref{sec:appendix_hyperparameters}.

\subsubsection{RAG Specific Training Loss Equations}
\label{sec:appendix_rag_loss_equations}
During the RAG specific fine-tuning stage, we optimize a joint loss function comprising generation, contrastive, and margin-based components: $\mathcal{L} = \mathcal{L}_{gen} + \mathcal{L}_{contrastive} + \mathcal{L}_{margin}$.

For the generation loss, let $\mathcal{D} = \{d_1, \dots, d_k\}$ be the uncompressed context documents and $\mathcal{C} = \{\repcontext^1, \dots, \repcontext^k\}$ be their corresponding compressed representations. The loss is the KL divergence between our model and the teacher reader:
$$
\mathcal{L}_{gen} = \sum_{t} D_{KL}(P_{\theta_{reader}}(a_t \mid a_{<t}, q, \mathcal{D}) \parallel P_{\theta}(a_t \mid a_{<t}, q, \mathcal{C}))
$$

For the retrieval representations $\repret$, we apply an InfoNCE contrastive loss over the positive document $d^+$ and all negatives $\mathcal{N}$ (hard negatives plus in-batch negatives), using a learned temperature $\tau$:
$$
\mathcal{L}_{contrastive} = - \log \frac{\exp(S(\repret^q, \repret^{d^+}) / \tau)}{\exp(S(\repret^q, \repret^{d^+}) / \tau) + \sum_{d^- \in \mathcal{N}} \exp(S(\repret^q, \repret^{d^-}) / \tau)}
$$

Finally, we distill the ranking capability of the teacher scoring model via a Margin MSE loss \citep{margin_mse} over the hard negatives $\mathcal{N}_{hard}$. We introduce a learnable scaling parameter $\alpha$ to account for score scale differences between the teacher and our student model:
$$
\mathcal{L}_{margin} = \frac{1}{|\mathcal{N}_{hard}|} \sum_{d^- \in \mathcal{N}_{hard}} (\Delta(S, d^-) - \alpha \Delta(T, d^-))^2
$$
where $\Delta(\phi, d^-) = \phi(q, d^+) - \phi(q, d^-)$ is the margin between the positive and negative document for a given scoring function $\phi$ (student $S$ or scoring teacher $T$).

\subsubsection{RAG Training Data Construction}
\label{sec:appendix_rag_training_data_construction}
For the NQ subset we use the data released by \citep{rlhn_hard_negatives} which includes a known positive passage, hard negative passages, and query-passage scores using Gemma 2 Reranker \citep{bge}.\footnote{https://huggingface.co/datasets/rlhn/default-680K-bge-reranker-v2-gemma} As this dataset lacks the answer text required for the generative component of our model, we augment the data with answers obtained from FlashRag \citep{flashrag}. To reduce the chance of false-negatives we remove any passage which has a higher reranker score than the known positive. For Trivia QA, we use the released training set, and use BM25 on the Wikipedia corpus to retrieve hard negatives.\footnote{https://huggingface.co/datasets/mandarjoshi/trivia\_qa/viewer/rc.wikipedia} We score the top 100 documents using the same Gemma 2 Reranker as the NQ data. As the Trivia QA data does not include a positive passage we select the passage which has the highest Gemma Reranker score and contains the answer string. We remove any other passages that also have the answer string to prevent potential false-negatives. As a final step, we remove any questions that does not contain at least eight negatives from both the NQ and Trivia QA subsets. The final dataset contains a combination of these subsets. The final training dataset contains around ~110k questions. We use a random (disjoint) subset of 1k questions to act as a validation set.

\subsubsection{RAG Weighted Negative Sampling}

\label{sec:appendix_rag_weighted_negative_sampling}
During RAG training we sample a negative $d^-_i$ with teacher score $s^-_i$ with probability:
\begin{equation}
    p_i = \frac{\exp(s^-_i  / \tau)}{\sum_{j=1}^{|\mathcal{N}|} \exp(s^-_j / \tau)},
\end{equation}
where $\mathcal{N}$ is the full set of negatives for a given query and $\tau$ is a hyperparameter which controls the temperature. Smaller values of $\tau$ mean the documents with higher teacher scores are selected more often while higher ones make the distribution closer to uniform.

\section{Additional Tables}
\input{tables/context_compression_same_arch}

\section{Discussion of Accuracy Drop Off with More Context Documents}
\label{sec:appendix_accuracy_drop_off}
As displayed in Figure \ref{fig:nq_k_scaling} when using the ECG model as the retriever both the ECG reader and standard RAG reader perform best with fewer documents, generally one or two, than the other retrieval models. In this section, we provide a theory for this behavior. Specifically, we believe it comes down to signal-to-noise ratio of the top documents and how this is impacted by retrieval quality.

To illustrate this, imagine a perfect retriever where the relevant document is always first. Additionally assume that each question also only has a single relevant document. Thus with a single context document the reader is guaranteed to have the necessary information to answer correctly, but as we expand to include more context documents we do not provide more useful information instead we add noise to the model's context. If we assume the reader will degrade to some extend with additional noise we can see how adding additional documents is actually harmful. We believe something similar is happening with the ECG model though to a lesser extent as it is not a perfect retriever. If for many of the questions the most relevant document is first additional documents are only going to dilute the relevant information, but if the retriever is weak and the first document does not contain the answer adding context documents improves the signal-to-noise ratio because there is no signal without a relevant document.

This theory is supported by the performance trends for ColBERT and BM25. ColBERT has the best performance around three documents with a gradual drop with additional documents, while BM25 shows pretty consistent improvement as more documents are added even up to five context documents. As BM25 is far weaker than ColBERT this supports the theory that more context documents are beneficial for worse retrievers.

This theory and the results of ECG reader using other retrievers demonstrates that the ECG reader is not limited to only leverage information from one relevant document, but when acting on a strong retriever's document ranking adding additional documents generally provides minimal useful information while often adding noise. We leave a more comprehensive study of this phenomenon for future works.

\section{ECG Model Ablations}
\label{sec:appendix_ecg_model_ablations}
To fully understand the impact of each component in our training pipeline we conduct a comprehensive set of ablation studies for the SmolLM ECG model. To minimize computational cost we used the same hyperparameters as the SmolLM ECG model except for the ablations that changed the pretraining stages where we did do a hyperparameter sweep but only for the pretraining stage. The results of the ablation are in Table \ref{tab:ecg_ablations} and the description and analysis of each ablation is below.
\begin{description}[leftmargin=8pt, labelindent=0pt]
    \item[Without Contrastive (Pretraining)] This ablation removes the in-batch contrastive loss during the self-supervised pretraining stage. The results show that while it is not strictly necessary for all retrieval tasks (as TriviaQA performance remains similar), it provides a notable performance boost on NQ, demonstrating that it is a useful addition for shaping the unified representations.
    
    \item[Without Generative Distillation] This replaces the knowledge distillation objective from the teacher reader with a standard next-token prediction loss during RAG fine-tuning. The resulting drop in performance highlights that generative distillation is crucial for producing high-quality generative outputs.
    
    \item[Without Pretraining] This skips the self-supervised pretraining phase entirely, moving straight to RAG specific training from the base language model. The decreased accuracy confirms that teaching the model to compress and understand context representations on unlabeled data is a meaningful factor in the ECG model's strong end-to-end performance.
    
    \item[Without Loss Scaling] This removes the dynamic loss scaling—specifically, the learned temperature and teacher score scale for the retrieval losses. The massive performance drop (falling to nearly half on NQ for $k=1$) indicates that without precise alignment of the teacher's scores to the student distribution, the multi-task objectives actively compete and degrade the shared representations.
    
    \item[Without Hard Negatives (Random)] This trains the model using uniformly random negatives instead of the weighted hard negatives selection used in the final ECG models. The performance gap demonstrates that leveraging harder negatives is a significant factor in developing robust retrieval capabilities.
    
    \item[With Negative Temperature 0.5 \& 0.25] These configurations alter the $\tau$ hyperparameter used for weighted negative sampling from the $0.15$ used for the standard models to $0.25$ and $0.5$. The severe degradation in performance under these constraints emphasizes that the model is highly sensitive to the exact hard-negative temperature, which dictates the difficulty of hard-negatives during training.
\end{description}

A major takeaway is that the contrastive loss during pretraining is not needed for all retrieval tasks as TriviaQA performance is similar with and without it. NQ has a more significant difference though which demonstrates it is still a useful addition. We also find that the generative distillation loss (as oppose to standard next token prediction) is important for high-quality generations. Doing self-supervised pretraining and leveraging harder negatives are also shown to be meaningful factors in the strong performance of the ECG models. Dynamic loss scaling, the learned temperature and teacher score scale for the retrieval losses, is a highly important factor; without it, the performance drops significantly, falling to around half (0.173 versus 0.343) for NQ when $k=1$. This indicates balancing the two losses and aligning the teacher's scores to the student distribution is crucial for high-quality representations. The exact hard-negative temperature seems to also be highly impactful, with a large degradation in performance for several configurations. Interestingly, we can see that as $k$ increases sometimes there is a large uptick in performance. We hypothesize that the model may have a consistent bias towards certain documents which usually appear in the top spots which pushes relevant information to lower $k$ values. Although the change from the negative temperature is extreme, we suspect this might indicate this parameter is more sensitive to hyperparameter choices and thus might do worse than it would do with tuned hyperparameters.

\begin{table*}[h]
\centering
\renewcommand{\arraystretch}{1.2}
\setlength{\tabcolsep}{6pt}

\resizebox{\textwidth}{!}{
\begin{tabular}{l @{\hskip 10pt} !{\color{rulecolor}\vrule} @{\hskip 10pt} ccccc @{\hskip 10pt} !{\color{rulecolor}\vrule} @{\hskip 10pt} ccccc}
\specialrule{\heavyrulewidth}{0pt}{0pt}

& \multicolumn{5}{c @{\hskip 10pt} !{\color{rulecolor}\vrule} @{\hskip 10pt}}{\textsc{NQ}} & \multicolumn{5}{c}{\textsc{TriviaQA}} \\
\cmidrule(lr{25pt}){2-6} \cmidrule(lr){7-11}

\textbf{Method} & \textbf{k=1} & \textbf{k=2} & \textbf{k=3} & \textbf{k=4} & \textbf{k=5} & \textbf{k=1} & \textbf{k=2} & \textbf{k=3} & \textbf{k=4} & \textbf{k=5} \\
\specialrule{\lightrulewidth}{0pt}{0pt}

\textbf{ECG} & \textbf{0.343} & \textbf{0.338} & \textbf{0.325} & \textbf{0.309} & \textbf{0.293} & \textbf{0.515} & \textbf{0.518} & \textbf{0.504} & \textbf{0.488} & \textbf{0.482} \\

\specialrule{\lightrulewidth}{0pt}{0pt}
\multicolumn{11}{>{\columncolor[gray]{0.95}}l}{\textbf{Ablations}} \\

\quad w/o Contrastive (Pretraining) & 0.337 & 0.335 & 0.325 & 0.313 & 0.295 & 0.518 & 0.516 & 0.503 & 0.488 & 0.477 \\
\quad w/o Generative Distillation   & 0.313 & 0.313 & 0.296 & 0.282 & 0.274 & 0.462 & 0.464 & 0.451 & 0.432 & 0.427 \\
\quad w/o Pretraining               & 0.300 & 0.302 & 0.293 & 0.276 & 0.262 & 0.470 & 0.480 & 0.469 & 0.450 & 0.430 \\
\quad w/o Loss Scaling              & 0.173 & 0.181 & 0.175 & 0.166 & 0.165 & 0.385 & 0.420 & 0.427 & 0.424 & 0.425 \\
\quad w/o Hard Negatives (Random)   & 0.308 & 0.320 & 0.314 & 0.301 & 0.302 & 0.514 & 0.518 & 0.507 & 0.493 & 0.493 \\
\quad w/ Negative Temp 0.5       & 0.325 & 0.334 & 0.318 & 0.308 & 0.305 & 0.051 & 0.065 & 0.083 & 0.442 & 0.478 \\
\quad w/ Negative Temp 0.25       & 0.077 & 0.219 & 0.308 & 0.317 & 0.324 & 0.042 & 0.031 & 0.028 & 0.025 & 0.022 \\

\specialrule{\heavyrulewidth}{0pt}{0pt}
\end{tabular}
}
\caption{Ablation study results for ECG on NQ and TriviaQA datasets across different $k$ values.}
\label{tab:ecg_ablations}
\end{table*}

\section{Hyperparameters}
\label{sec:appendix_hyperparameters}
This section contains the hyperparameters used for each ECG and baseline model trained (see Tables \ref{tab:autoencoder_context_compression_hparams}, \ref{tab:autoencoder_ecg_hparams}, \ref{tab:reader_hparams}, \ref{tab:context_comp_hparams}, \ref{tab:ecg_hparams}, \ref{tab:parametric_hparams}, and \ref{tab:ssl_hparams}). Additionally, we include the information on the hyperparameter sweep we conducted for the ECG models and baseline models in Table \ref{tab:full_sweep_values}.

\begin{table}[H]
    \centering

    \begin{tabular}{lccc}
        \toprule
        \textbf{Hyperparameter} & \textbf{SmolLM-v2 (w/ MLPs)} & \textbf{SmolLM-v2} & \textbf{Gemma 3} \\
        \midrule
        Includes MLPs & \ding{51} & \ding{55} & \ding{55} \\
        Single Batch Size & 16 & 16 & 6 \\
        Gradient Accumulation & 1 & 1 & 2 \\
        Learning Rate & 3e-4 & 3e-4 & 2e-5 \\
        Weight Decay & 0.0 & 0.0 & 0.0 \\
        Train Epochs & 1 & 1 & 1 \\
        \bottomrule
    \end{tabular}
    \caption{Hyperparameters for Self-Supervised Pretraining stage for models without contrastive loss (i.e. with only the generative loss). These models were used to initialize the context compression baseline models including the SmolLM model with the same architecture as the ECG model. The SmolLM model trained with MLPs was also used for the ECG ablation which does not use contrastive loss for pretraining.}
    \label{tab:autoencoder_context_compression_hparams}
\end{table}

\begin{table}[H]
    \centering

    \begin{tabular}{lcc}
        \toprule
        \textbf{Hyperparameter} & \textbf{SmolLM-v2} & \textbf{Gemma 3} \\
        \midrule
        Single Batch Size & 16 & 6 \\
        Gradient Accumulation & 1 & 2 \\
        Learning Rate & 4e-4 & 9e-5 \\
        Weight Decay & 0.0 & 0.0 \\
        Train Epochs & 1 & 1 \\
        \bottomrule
    \end{tabular}
    \caption{Hyperparameters for Self-Supervised Pretraining stage for models with contrastive loss. These models were used as the starting checkpoints for RAG training for the ECG models including the ablation ECG models if they did not use a different pretraining strategy.}
    \label{tab:autoencoder_ecg_hparams}
\end{table}

\begin{table}[H]
    \centering
    \begin{tabular}{lcc}
        \toprule
        \textbf{Hyperparameter} & \textbf{SmolLM-v2} & \textbf{Gemma 3} \\
        \midrule
        Single Batch Size & 16 & 8 \\
        Gradient Accumulation & 1 & 2 \\
        Learning Rate & 1e-4 & 5e-6 \\
        Weight Decay & 0.1 & 0.1 \\
        Train Epochs & 10 & 10 \\
        \bottomrule
    \end{tabular}
    \caption{Hyperparameters for RAG training for the baseline Reader Models.}
    \label{tab:reader_hparams}
\end{table}

\begin{table}[H]
    \centering
    \begin{tabular}{lccc}
        \toprule
        \textbf{Hyperparameter} & \textbf{SmolLM-v2 (w/ MLPs)} & \textbf{SmolLM-v2} & \textbf{Gemma 3} \\
        \midrule
        Includes MLPs & \ding{51} & \ding{55} & \ding{55} \\
        Single Batch Size & 8 & 8 & 3 \\
        Gradient Accumulation & 2 & 2 & 4 \\
        Learning Rate & 3e-4 &  3e-4 & 1e-4 \\
        Weight Decay &  0.15 & 0.05 & 0.05 \\
        Train Epochs & 20 & 20 & 20 \\
        \bottomrule
    \end{tabular}
    \caption{Hyperparameters for RAG training for the baseline context compression models.}
    \label{tab:context_comp_hparams}
\end{table}

\begin{table}[H]
    \centering
    \begin{tabular}{lcc}
        \toprule
        \textbf{Hyperparameter} & \textbf{SmolLM-v2} & \textbf{Gemma 3} \\
        \midrule
        Single Batch Size & 16 & 6 \\
        Gradient Accumulation & 1 & 2 \\
        Learning Rate & 3e-4 & 9e-5 \\
        Weight Decay & 0.05 & 0.0 \\
        Train Epochs & 20 & 20 \\
        \bottomrule
    \end{tabular}
    \caption{Hyperparameters for RAG training for the ECG models. For the ablation models we use the same hyperparameters.}
    \label{tab:ecg_hparams}
\end{table}

\begin{table}[H]
    \centering
    \begin{tabular}{lcc}
        \toprule
        \textbf{Hyperparameter} & \textbf{SmolLM-v2} & \textbf{Gemma 3} \\
        \midrule
        Single Batch Size & 16 & 8 \\
        Gradient Accumulation & 1 & 2 \\
        Learning Rate & 8e-5 & 7e-6 \\
        Weight Decay & 0.2 & 0.0 \\
        \bottomrule
    \end{tabular}
    \caption{Hyperparameters for RAG training for the baseline parametric models.}
    \label{tab:parametric_hparams}
\end{table}

\begin{table}[H]
    \centering
    \begin{tabular}{lcc}
        \toprule
        \textbf{Hyperparameter} & \textbf{SmolLM-v2} & \textbf{Gemma 3} \\
        \midrule
        Optimizer            & AdamW & AdamW \\
        Scheduler            & Linear & Linear \\
        Warmup Ratio         & 0.05 & 0.05 \\
        Negative Temperature & 0.15 & 0.15 \\
        \# GPUs              & 8 & 8 \\
        \bottomrule
    \end{tabular}
    \caption{Shared hyperparameters for all training runs.}
    \label{tab:ssl_hparams}
\end{table}

\begin{table}[H]
    \centering

    \footnotesize
    \resizebox{\textwidth}{!}{
    \begin{tabular}{lp{8.5cm}l}
        \toprule
        \textbf{Model Category} & \textbf{Learning Rates (LR) Explored} & \textbf{Weight Decay (WD)} \\
        \midrule
        Pretraining (All) & 7e-4, 6e-4, 5e-4, 4e-4, 3e-4, 2e-4, 1e-4, 9e-5, 8e-5, 7e-5, 6e-5, 5e-5, 4e-5, 3e-5, 2e-5, 1e-5, 9e-6, 3e-6 & 0.0, 0.1, 0.2 \\
        \midrule
        Reader & 4e-4, 3e-4, 2e-4, 1e-4, 9e-5, 8e-5, 7e-5, 6e-5, 5e-5, 4e-5, 3e-5, 2e-5, 1e-5, 9e-6, 8e-6, 7e-6, 6e-6, 5e-6, 4e-6, 3e-6, 2e-6, 1e-6, 9e-7 & 0.0, 0.1, 0.2 \\
        \midrule
        Parametric & 4e-4, 3e-4, 2e-4, 1e-4, 9e-5, 8e-5, 7e-5, 6e-5, 5e-5, 4e-5, 3e-5, 2e-5, 1e-5, 9e-6, 8e-6, 7e-6, 6e-6 & 0.0, 0.1, 0.2 \\
        \midrule
        Context Comp. & 4e-4, 3e-4, 2e-4, 1e-4, 9e-5, 8e-5, 7e-5, 6e-5, 5e-5, 4e-5 & 0.0, 0.05, 0.1, 0.15, 0.2 \\
        \midrule
        \textbf{ECG} & 6e-4, 5e-4, 4e-4, 3e-4, 2e-4, 1e-4, 9e-5, 8e-5, 7e-5, 6e-5, 5e-5, 4e-5, 3e-5, 2e-5 & 0.0, 0.05, 0.1, 0.15, 0.2 \\
        \bottomrule
    \end{tabular}
    }
    \caption{Hyperparameter Search Space. We performed a grid search over the union of learning rates (LR) and weight decay (WD) values explored for both SmolLM-v2 and Gemma 3 architectures. Note that for each model we started with an expected reasonable learning rate based on initial experiments and continued to search until the best LR was not the largest or smallest LR, thus each model was usually not evaluated on each LR value. This is also why some models have substantially different starting and ending points.}
    \label{tab:full_sweep_values}
\end{table}

\section{Prompts}

\begin{figure*}[h]
\centering
\begin{prompt}{Reader Model Prompt}
Generate a response to the following question using the context.

Question: \{question\}

Context: \{context text\}

Response:
\end{prompt}
\caption{Prompt used for standard RAG reader model. Note that context text was separated by two new lines.}
\end{figure*}

\begin{figure*}[h]
\centering
\begin{prompt}{Parametric Model Prompt}
Generate a response to the following question.

Question: \{question\}

Response:
\end{prompt}
\caption{Prompt used for parametric (no retrieval) model.}
\end{figure*}

\begin{figure*}[h]
\centering
\begin{prompt}{Encoding/Compression Prompt}
Encode the following text into embedding tokens. \{text\} \texttt{<emb\_start>} \texttt{<emb>} $\cdots$ \texttt{<emb>} \texttt{<emb\_stop>}
\end{prompt}
\caption{Prompt used for encoding/compression for the ECG and context compression models.}
\end{figure*}

\begin{figure*}[h]
\centering
\begin{prompt}{Generation Prompt from Compressed Context}
Generate a response to the following question using the embedded context.

Question: \{question\}

Context: \{compressed context vectors\}

Response:
\end{prompt}
\caption{Prompt used for generating with compressed context representations. This is used for both the context compression and ECG models.}
\end{figure*}

\end{document}

%% file: figures/main_ecg_overview_figure.tex
\begin{figure*}[tbp]
    \centering
    \resizebox{\textwidth}{!}{
    \begin{tikzpicture}[
        scale=0.85, 
        transform shape,
        font=\sffamily,
        >=Stealth,
        /utils/exec={\newcommand{\colOFF}{0}    
                     \newcommand{\colRET}{5.5}  
                     \newcommand{\colCOM}{10.8}  
                     \newcommand{\colGEN}{12.5} 
        },
        node distance=0.5cm and 0.5cm,
        process/.style={rectangle, rounded corners=4pt, minimum width=2.2cm, minimum height=0.7cm, align=center, draw=blue!50!black, fill=blue!5, thick},
        model_block/.style={process, fill=violet!10, draw=violet!60!black, align=center},
        sub_block/.style={process, minimum width=1.5cm, minimum height=0.6cm, fill=teal!10, draw=teal!60!black, font=\small\bfseries, align=center},
        data/.style={rectangle, draw=green!60!black, fill=green!5, dashed, thick, align=center, minimum width=1.8cm, minimum height=0.7cm},
        token_base/.style={anchor=center, inner sep=2pt},
        text_token/.style={token_base, font=\normalsize\ttfamily}, 
        special_token/.style={token_base, font=\small\ttfamily},
        vector/.style={rectangle, draw=orange!80!black, fill=orange!30, minimum width=0.2cm, minimum height=0.4cm, inner sep=0pt, anchor=center},
        context_vector/.style={vector, draw=purple!80!black, fill=purple!20},
        storage/.style={cylinder, shape border rotate=90, aspect=0.25, draw=black, fill=gray!20, minimum width=1.5cm, minimum height=0.9cm, align=center},
        dots/.style={font=\large, anchor=center}
    ]
    
    \node[text_token] (d_t) at (\colOFF, 0) {$t_{1:n}$};
    \node[special_token, right=0.1cm of d_t] (d_es) {<emb\_s>};
    \node[special_token, right=0.1cm of d_es] (d_e) {<emb>$_{1:n}$};
    \node[special_token, right=0.1cm of d_e] (d_ee) {<emb\_e>};
    
    \node[fit=(d_t) (d_ee), draw=gray, dashed, rounded corners, label=above:\small Embedding] (doc_input) {};
    
    \node[model_block, below=0.6cm of doc_input] (lm_enc) {$\theta_{LM}$};
    \node[sub_block, below=0.5cm of lm_enc] (theta_ret_doc) {$\theta_{ret}$};
    
    \node[vector, below=0.6cm of theta_ret_doc, xshift=-0.12cm] (ret_v1) {};
    \node[vector, right=0.08cm of ret_v1] (ret_vn) {};
    \node[right=0.05cm of ret_vn, color=orange!70!black, font=\scriptsize, align=left] {Document\\Embeddings};
    \node[fit=(ret_v1) (ret_vn)] (ret_output) {};

    \node[storage, below=0.45cm of ret_output] (index) {\textbf{Index}};
    
    \draw[->, thick] (doc_input.south) -- (lm_enc.north);
    \draw[->, thick] (lm_enc.south) -- (theta_ret_doc.north);
    \draw[->, thick] (theta_ret_doc.south) -- (ret_output.north);
    \draw[->, thick] (ret_output.south) -- (index.north);
    
    \node[above=0.5cm of doc_input, font=\bfseries, color=black] {OFFLINE};
    
    \node[text_token] (q_t) at (\colRET, 0) {$q_{1:m}$};
    \node[special_token, right=0.1cm of q_t] (q_es) {<emb\_s>};
    \node[special_token, right=0.1cm of q_es] (q_e) {<emb>$_{1:m}$};
    \node[special_token, right=0.1cm of q_e] (q_ee) {<emb\_e>};

    \node[fit=(q_t) (q_ee), draw=gray, dashed, rounded corners, label=above:\small Retrieval] (query_input) {};
    \node[model_block, below=0.6cm of query_input] (lm_query) {$\theta_{LM}$};
    \node[sub_block, below=0.5cm of lm_query] (theta_ret_query) {$\theta_{ret}$};
    
    \node[vector, below=0.6cm of theta_ret_query, xshift=-0.12cm] (q_ret_v1) {};
    \node[vector, right=0.08cm of q_ret_v1] (q_ret_vn) {};
    \node[right=0.05cm of q_ret_vn, color=orange!70!black, font=\scriptsize, align=left] {Query\\Embeddings};
    \node[fit=(q_ret_v1) (q_ret_vn)] (q_ret_output) {};
    
    \node[process, fill=red!5, draw=red!60!black, below=0.6cm of q_ret_output] (retrieval_op) {MaxSim Search};
    
    \draw[->, thick] (query_input.south) -- (lm_query.north);
    \draw[->, thick] (lm_query.south) -- (theta_ret_query.north);
    \draw[->, thick] (theta_ret_query.south) -- (q_ret_output.north);
    \draw[->, thick] (q_ret_output.south) -- (retrieval_op.north);
    \draw[->, thick, dashed] (index.east |- retrieval_op.west) -- (retrieval_op.west);

    \node[vector] (tk_mid_v) at (\colCOM, -5.0) {}; 
    \node[vector, left=0.1cm of tk_mid_v] (tk_d1_v1) {};
    \node[vector, right=0.1cm of tk_mid_v] (tk_d1_vn) {};
    \node[dots, above=0.05cm of tk_mid_v, scale=0.6] (tk_v_dots) {$\vdots$};
    \node[vector, above=0.1cm of tk_v_dots] (tk_dk_v_mid) {};
    \node[vector, left=0.1cm of tk_dk_v_mid] (tk_dk_v1) {};
    \node[vector, right=0.1cm of tk_dk_v_mid] (tk_dk_vn) {};
    \node[fit=(tk_d1_v1) (tk_dk_vn) (tk_dk_v_mid)] (top_k_box) {};
    
    \node[sub_block, above=0.6cm of top_k_box] (theta_comp) {$\theta_{comp}$};
    
    \node[context_vector, above=0.6cm of theta_comp] (ctx_mid_v) {};
    \node[context_vector, left=0.1cm of ctx_mid_v] (ctx_d1_v1) {};
    \node[context_vector, right=0.1cm of ctx_mid_v] (ctx_d1_vn) {};
    \node[dots, above=0.05cm of ctx_mid_v, scale=0.6] (ctx_v_dots) {$\vdots$};
    \node[context_vector, above=0.1cm of ctx_v_dots] (ctx_dk_v_mid) {};
    \node[context_vector, left=0.1cm of ctx_dk_v_mid] (ctx_dk_v1) {};
    \node[context_vector, right=0.1cm of ctx_dk_v_mid] (ctx_dk_vn) {};
    \node[fit=(ctx_d1_v1) (ctx_dk_vn)] (context_reps_box) {};

    \draw[->, thick] (retrieval_op.east) -- (top_k_box.west);
    \draw[->, thick] (top_k_box.north) -- (theta_comp.south);
    \draw[->, thick] (theta_comp.north) -- (context_reps_box.south);

    \node[text_token] (g_q) at (\colGEN, 0) {$q_{1:m}$};
    \node[dots, right=0.1cm of g_q, scale=0.7] (g_sep) {..};
    \node[special_token, right=0.1cm of g_sep] (g_es1) {<emb\_s>};
    \node[context_vector, right=0.1cm of g_es1] (g_cv1) {};
    \node[dots, right=0.05cm of g_cv1, scale=0.7] (g_cdots) {..};
    \node[context_vector, right=0.05cm of g_cdots] (g_cvn) {};
    \node[special_token, right=0.1cm of g_cvn] (g_ee1) {<emb\_e>};

    \node[fit=(g_q) (g_ee1), draw=gray, dashed, rounded corners, inner sep=4pt, label=above:\small Generation] (gen_input_box) {};

    \node[model_block, below=1.2cm of gen_input_box] (lm_gen) {$\theta_{LM}$};
    \node[data, below=0.6cm of lm_gen] (answer) {Final Answer};
    
    \draw[->, thick] (gen_input_box.south) -- (lm_gen.north);
    \draw[->, thick] (lm_gen.south) -- (answer.north);

    \draw[->, thick] (context_reps_box.north) |- (gen_input_box.west);

    \node[above=0.5cm of query_input, font=\bfseries, color=black, xshift=3.5cm] {ONLINE};

    \begin{scope}[on background layer]
        \coordinate (top_mid) at ($(doc_input.north east)!0.5!(query_input.north west)$);
        \coordinate (bottom_mid) at ($(index.south east)!0.5!(retrieval_op.south west)$);
        
        \draw[dashed, ultra thick, gray!20] 
            ([yshift=0.5cm]top_mid) -- ([yshift=-0.5cm]top_mid |- bottom_mid);

        \node[fit=(tk_d1_v1) (tk_dk_vn), draw=orange!60!black, align=left, fill=orange!5, dashed, rounded corners, label={[font=\scriptsize, color=orange!70!black, align=left]right:Top-$k$ Document\\Embeddings}] {};
        \node[fit=(ctx_d1_v1) (ctx_dk_vn), draw=purple!60!black, fill=purple!5, dashed, rounded corners, label={[font=\scriptsize, color=purple!70!black, align=left]right:Top-k Compressed\\Documents}] {};
    \end{scope}

    \end{tikzpicture}
    } %
    \caption{Inference usage of the ECG (Embed, Compress, Generate) models.}
    \label{fig:inference_usage_of_ecg_model}
\end{figure*}
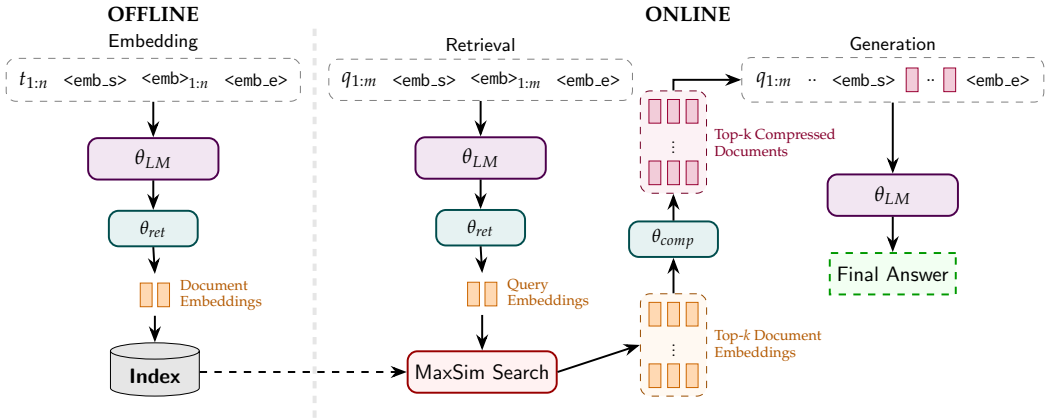

%% file: tables/main_results_fixed_context_budget.tex
\definecolor{rulecolor}{gray}{0.8}
\begin{table*}[!t]
\centering
\renewcommand{\arraystretch}{1.2}
\setlength{\tabcolsep}{10pt}

\resizebox{\textwidth}{!}{
\begin{tabular}{l @{\hskip 10pt} !{\color{rulecolor}\vrule} @{\hskip 10pt} ccc @{\hskip 10pt} !{\color{rulecolor}\vrule} @{\hskip 10pt} ccc}
\specialrule{\heavyrulewidth}{0pt}{0pt}

& \multicolumn{3}{c @{\hskip 10pt} !{\color{rulecolor}\vrule} @{\hskip 10pt}}{\textsc{SmolLM 2 135M}} & \multicolumn{3}{c}{\textsc{Gemma 3 1B}} \\
\cmidrule(lr{25pt}){2-4} \cmidrule(lr){5-7}

\textbf{Method} & \textbf{Context Budget} & \textbf{NQ} & \textbf{Trivia QA} & \textbf{Context Budget} & \textbf{NQ} & \textbf{Trivia QA} \\
\specialrule{\lightrulewidth}{0pt}{0pt}

\multicolumn{7}{>{\columncolor[gray]{0.95}}l}{\textbf{Baselines}} \\

Parametric (No Ret.) & - & 0.069 & 0.110 & - & 0.142 & 0.267 \\
\textit{RAG:} & & & & & & \\
\quad BM25 & 32 & 0.047 & 0.195 & 16 & 0.061 & 0.241 \\
\quad Dense & 32 & 0.096 & 0.231 & 16 & 0.099 & 0.242 \\
\quad ColBERT & 32 & 0.106 & 0.297 & 16 & 0.104 & 0.303 \\
\textit{RAG w/ Context Comp:} & & & & & & \\
\quad BM25 & 32 & 0.087 & 0.242 & 16 & 0.144 & 0.318 \\
\quad Dense & 32 & 0.192 & 0.303 & 16 & 0.245 & 0.351 \\
\quad ColBERT & 32 & 0.188 & 0.351 & 16 & 0.247 & 0.393 \\

\specialrule{\lightrulewidth}{0pt}{0pt}
\multicolumn{7}{>{\columncolor[gray]{0.95}}l}{\textbf{Ours}} \\
\textbf{ECG} & 32 & \textbf{0.343} & \textbf{0.515} & 16 & \textbf{0.361} & \textbf{0.540} \\
\specialrule{\heavyrulewidth}{0pt}{0pt}
\end{tabular}
}
\caption{Comparison between Exact Match (EM) accuracy for ECG models (ours) and baseline models with a fixed context budget, which is the number of vectors used to represent the context documents.}
\label{tab:main_results_fixed_context_budget}
\end{table*}

%% file: tables/main_results_fixed_performance.tex
\begin{table}[!t]

\centering
\resizebox{\textwidth}{!}{
\small 
\setlength{\tabcolsep}{3pt}
\setlength{\aboverulesep}{0pt}
\setlength{\belowrulesep}{0pt}

\renewcommand{\arraystretch}{1.5} %

\definecolor{rulecolor}{gray}{0.8}

\begin{tabular}{l c @{\hskip 10pt} !{\color{rulecolor}\vrule} @{\hskip 10pt} ccc ccc @{\hskip 10pt} !{\color{rulecolor}\vrule} @{\hskip 10pt} ccc ccc}

\specialrule{\heavyrulewidth}{0pt}{0pt}

& & \multicolumn{6}{c @{\hskip 10pt} !{\color{rulecolor}\vrule} @{\hskip 10pt}}{\textsc{SmolLM 2 135M}} & \multicolumn{6}{c}{\textsc{Gemma 3 1B}} \\

\cmidrule(lr{25pt}){3-8} \cmidrule(lr){9-14}

& & \multicolumn{3}{c}{NQ} & \multicolumn{3}{c @{\hskip 10pt} !{\color{rulecolor}\vrule} @{\hskip 10pt}}{Trivia QA} & \multicolumn{3}{c}{NQ} & \multicolumn{3}{c}{Trivia QA} \\

\cmidrule(lr){3-5} \cmidrule(lr{25pt}){6-8} \cmidrule(lr){9-11} \cmidrule(lr){12-14}

\textbf{Method} & \textbf{Disk Space / 10k} & \textbf{Context} & \textbf{Comp.} & \textbf{EM} & \textbf{Context} & \textbf{Comp.} & \textbf{EM} & \textbf{Context} & \textbf{Comp.} & \textbf{EM} & \textbf{Context} & \textbf{Comp.} & \textbf{EM} \\

\specialrule{\lightrulewidth}{0pt}{0pt}
\multicolumn{14}{>{\columncolor[gray]{0.95}}l}{\textbf{Baselines}} \\

Parametric (No Ret.) & 0.0 & - & - & 0.069 & - & - & 0.110 & - & - & 0.142 & - & - & 0.267 \\
\textit{RAG:} & & & & & & & & & & & & & \\
\quad BM25 & $0.0^*$ & 256 & $8\times$ & 0.163 & 256 & $8\times$ & 0.409 & 256 & $16\times$ & 0.195 & 256 & $16\times$ & 0.468 \\
\quad Dense & $0.031$ & 256 & $8\times$ & 0.327 & 256 & $8\times$ & 0.477 & 256 & $16\times$ & 0.358 & 256 & $16\times$ & 0.521 \\
\quad ColBERT & $0.744$ & 256 & $8\times$ & 0.328 & 160 & $5\times$ & \textbf{0.520} & 224 & $14\times$ & \textbf{0.361} & 128 & $8\times$ & \textbf{0.545} \\
\multicolumn{2}{l}{\textit{RAG w/ Context Comp:}} & & & & & & & & & & & & \\
\quad BM25 & $0.737$ & 96 & $3\times$ & 0.102 & 96 & $3\times$ & 0.278 & 64 & $4\times$ & 0.177 & 80 & $5\times$ & 0.370 \\
\quad Dense & $0.768$ & 96 & $3\times$ & 0.202 & 96 & $3\times$ & 0.324 & 64 & $4\times$ & 0.265 & 80 & $5\times$ & 0.392 \\
\quad ColBERT & $1.481$ & 96 & $3\times$ & 0.217 & 96 & $3\times$ & 0.372 & 64 & $4\times$ & 0.278 & 80 & $5\times$ & 0.438 \\
\specialrule{\lightrulewidth}{0pt}{0pt}

\multicolumn{14}{>{\columncolor[gray]{0.95}}l}{\textbf{Ours}} \\
\textbf{ECG} & $0.737$ & \textbf{32} & - & \textbf{0.343} & \textbf{32} & - & 0.515 & \textbf{16} & - & \textbf{0.361} & \textbf{16} & - & 0.540 \\
\specialrule{\heavyrulewidth}{0pt}{0pt}
\end{tabular}
}
\caption{Performance comparison of the Exact Match (EM) accuracy between the ECG models and baseline models at a fixed performance level. Disk Space / 10k, represents the size the index/embeddings for 10k passages. For all the embedding based retrieval models and ECG models we only count the size of the embeddings as otherwise the size would be heavily dependent on the indexing approach used which is orthogonal to our contribution. The columns Context and Comp. represent the context budget and relative compression of the ECG model. *The size of BM25 is actually, $9.23\mathrm{e}{-6}$, but for consistent formatting we rounded it to $0.0$ in the table.}
\label{tab:main_fixed_performance}
\end{table}

%% file: figures/comparison_ecg_reader_retrieval.tex
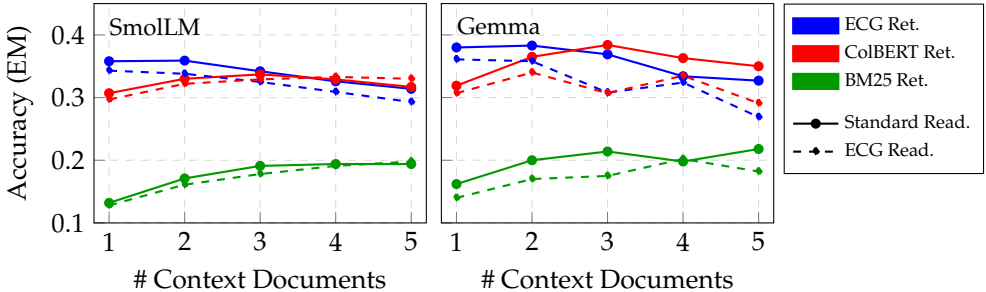
\begin{figure}[tbp]
\centering
\begin{tikzpicture}
\begin{groupplot}[
    group style={
        group size=2 by 1,
        horizontal sep=0.2cm,
        y descriptions at=edge left %
    },
    width=6cm, %
    height=4.5cm,
    xlabel={\# Context Documents},
    xmin=0.8, xmax=5.2,
    ymin=0.1, ymax=0.45,
    xtick={1,2,3,4,5},
    grid=both,
    grid style={dashed, gray!30},
    xtick pos=bottom, %
    ytick pos=left,   %
    every axis plot/.append style={thick, mark size=1.5pt},
]

\nextgroupplot[ylabel={Accuracy (EM)}]

\node[anchor=north west] at (rel axis cs: 0.02, 0.98) {\footnotesize SmolLM};

\addplot[color=blue, mark=*, forget plot] coordinates {
    (1, 0.358) (2, 0.359) (3, 0.342) (4, 0.326) (5, 0.314)
};
\addplot[color=red, mark=*, forget plot] coordinates {
    (1, 0.307) (2, 0.330) (3, 0.337) (4, 0.329) (5, 0.317)
};
\addplot[color=green!60!black, mark=*, forget plot] coordinates {
    (1, 0.132) (2, 0.171) (3, 0.191) (4, 0.194) (5, 0.194)
};
\addplot[color=blue, mark=diamond*, dashed, forget plot] coordinates {
    (1, 0.343) (2, 0.338) (3, 0.325) (4, 0.309) (5, 0.293)
};
\addplot[color=red, mark=diamond*, dashed, forget plot] coordinates {
    (1, 0.297) (2, 0.322) (3, 0.329) (4, 0.333) (5, 0.330)
};
\addplot[color=green!60!black, mark=diamond*, dashed, forget plot] coordinates {
    (1, 0.128) (2, 0.161) (3, 0.178) (4, 0.191) (5, 0.198)
};

\nextgroupplot[
    legend pos=outer north east,
    legend columns=1,
    legend style={font=\scriptsize, cells={anchor=west}, inner sep=3pt}
]

\node[anchor=north west] at (rel axis cs: 0.02, 0.98) {\footnotesize Gemma};

\addplot[color=blue, mark=*, forget plot] coordinates {
    (1, 0.380) (2, 0.383) (3, 0.369) (4, 0.334) (5, 0.327)
};
\addplot[color=red, mark=*, forget plot] coordinates {
    (1, 0.319) (2, 0.365) (3, 0.384) (4, 0.363) (5, 0.350)
};
\addplot[color=green!60!black, mark=*, forget plot] coordinates {
    (1, 0.162) (2, 0.200) (3, 0.214) (4, 0.198) (5, 0.218)
};
\addplot[color=blue, mark=diamond*, dashed, forget plot] coordinates {
    (1, 0.361) (2, 0.358) (3, 0.308) (4, 0.324) (5, 0.269)
};
\addplot[color=red, mark=diamond*, dashed, forget plot] coordinates {
    (1, 0.307) (2, 0.340) (3, 0.307) (4, 0.335) (5, 0.291)
};
\addplot[color=green!60!black, mark=diamond*, dashed, forget plot] coordinates {
    (1, 0.140) (2, 0.170) (3, 0.175) (4, 0.202) (5, 0.182)
};

\addlegendimage{area legend, fill=blue, draw=blue}
\addlegendentry{ECG Ret.}

\addlegendimage{area legend, fill=red, draw=red}
\addlegendentry{ColBERT Ret.}

\addlegendimage{area legend, fill=green!60!black, draw=green!60!black}
\addlegendentry{BM25 Ret.}

\addlegendimage{empty legend}
\addlegendentry{}

\addlegendimage{color=black, mark=*, solid, thick}
\addlegendentry{Standard Read.}

\addlegendimage{color=black, mark=diamond*, dashed, thick}
\addlegendentry{ECG Read.}

\end{groupplot}
\end{tikzpicture}
\caption{Accuracy (EM) of various retriever and reader combinations on NQ across different values of $k$ for SmolLM and Gemma models.}
\label{fig:nq_k_scaling}
\end{figure}

%% file: tables/context_compression_same_arch.tex
\begin{table}[H]

\centering
\resizebox{\textwidth}{!}{
\small 
\setlength{\tabcolsep}{3pt}
\setlength{\aboverulesep}{0pt}
\setlength{\belowrulesep}{0pt}

\renewcommand{\arraystretch}{1.5} %

\definecolor{rulecolor}{gray}{0.8}

\begin{tabular}{l @{\hskip 10pt} !{\color{rulecolor}\vrule} @{\hskip 10pt} c @{\hskip 10pt} !{\color{rulecolor}\vrule} @{\hskip 10pt} cc @{\hskip 10pt} !{\color{rulecolor}\vrule} @{\hskip 10pt} c @{\hskip 10pt} !{\color{rulecolor}\vrule} @{\hskip 10pt} cc}

\specialrule{\heavyrulewidth}{0pt}{0pt}

& \multicolumn{3}{c @{\hskip 10pt} !{\color{rulecolor}\vrule} @{\hskip 10pt}}{NQ} & \multicolumn{3}{c}{Trivia QA} \\

\cmidrule(lr{25pt}){2-4} \cmidrule(lr){5-7}

\textbf{Method} & \textbf{EM ($k=1$)} & \textbf{EM ($k=k'$)} & \textbf{$k'$} & \textbf{EM ($k=1$)} & \textbf{EM ($k=k'$)} & \textbf{$k'$} \\

\specialrule{\lightrulewidth}{0pt}{0pt}
\multicolumn{7}{>{\columncolor[gray]{0.95}}l}{\textbf{Baselines}} \\

Compression & 0.188 & 0.217 & 3 & 0.351 & 0.372 & 3 \\
Compression w/ Same Arch & 0.224 & 0.259 & 4 & 0.386 & 0.421 & 3 \\

\specialrule{\lightrulewidth}{0pt}{0pt}
\multicolumn{7}{>{\columncolor[gray]{0.95}}l}{\textbf{Ours}} \\

ECG - Reader & 0.297 & 0.333 & 4 & 0.485 & \textbf{0.518} & 3 \\
\specialrule{\lightrulewidth}{0pt}{0pt}

\textbf{ECG} & \textbf{0.343} & \textbf{0.343} & 1 & \textbf{0.515} & \textbf{0.518} & 2 \\

\specialrule{\heavyrulewidth}{0pt}{0pt}
\end{tabular}
}
\caption{Performance comparison between context compression baseline and a modified context compression model with the same architecture as the ECG models using ColBERT retrievals. ECG-Reader is the ECG model used only as a reader on top of the ColBERT retrievals. We show the EM for a single document and the best EM found for documents (1-5) where the number of documents is given as $k'$. }
\label{tab:context_compression_same_arch}
\end{table}